\documentclass{ws-procs9x6}

\newcommand{\ket}[1]{|{#1}\rangle}

\begin{document}

\title{QUANTUM FIELDS, DARK MATTER AND NON-STANDARD WIGNER CLASSES}

\author{A. B. GILLARD}

\address{Physics Department, University of Canterbury,\\
Private Bag 4800, Christchurch 8140, New Zealand\\
E-mail: abg22@student.canterbury.ac.nz}

\author{ and B. M. S. MARTIN}

\address{Mathematics \& Statistics Department, University of Canterbury,\\
Private Bag 4800, Christchurch 8140, New Zealand\\
E-mail: B.Martin@math.canterbury.ac.nz}

\begin{abstract}
 The Elko field of Ahluwalia and Grumiller is a quantum field for massive spin-$1/2$ particles.  It has been suggested as a candidate for dark matter.  We discuss our attempts to interpret the Elko field as a quantum field in the sense of Weinberg.  Our work suggests that one should investigate quantum fields based on representations of the full Poincar\'e group which belong to one of the non-standard Wigner classes. 
\end{abstract}

\keywords{quantum field; Elko field; dark matter; non-standard Wigner class}

\bodymatter

\section{Introduction}\label{gm:intro}

In 2005 Ahluwalia and Grumiller introduced a new quantum field for massive spin-$1/2$ particles, which they called the {\em Elko} field\cite{gm:AhlGru05,gm:AhlGru05a}.
They proposed the Elko field as a candidate for dark matter.
It is natural to ask how the Elko field fits into Weinberg's formulation of quantum field theory (Ref.~\refcite{gm:Weinberg95}, Ch.~5).
In this note we report on our recent investigations into this question.  Forthcoming work\cite{gm:AhlLeeSch08} of Ahluwalia, Lee and Schritt
also deals with aspects of this and related questions.

We begin by recalling the Elko field and its properties (\sref{gm:sec:elkos}).  In \sref{gm:sec:weinberg} and \sref{gm:sec:DiracvsElko} we briefly describe Weinberg's construction of quantum fields and compare the Elko field with the Dirac field.  In the final section, \sref{gm:sec:nonstdclasses}, we discuss directions for future research.

\section{Review of Elko Fields}\label{gm:sec:elkos}

We start with some notation.  We denote the strict Lorentz and Poincar\'e groups by ${\mathcal L}_0$ and ${\mathcal P}_0$ respectively, and the full Lorentz and Poincar\'e groups --- which include space inversion ${\sf P}$ and time reversal ${\sf T}$ --- by ${\mathcal L}$ and ${\mathcal P}$ respectively.  We represent elements of ${\mathcal L}_0$ by the symbol $\Lambda$, and elements of ${\mathcal P}_0$ by pairs $(\Lambda,a)$, where $a$ is a space-time translation.

The Elko field\footnote{We have absorbed a $p$-dependent factor in the integrand into the definition of the spinors $\lambda({\mathbf p},\beta)$, and we have done the same for the Dirac field below.} is given by
\begin{equation}\label{gm:eqn:elko}
 \eta_i(x)= \int {\rm d}^3{\mathbf p} \sum_\beta \left[e^{-ip_\mu x^\mu}\lambda_i^{\rm S}({\mathbf p},\beta)c_\beta({\mathbf p}) + e^{+ip_\mu x^\mu}\lambda_i^{\rm A}({\mathbf p},\beta)d^\dagger_\beta({\mathbf p})\right],
\end{equation}
where the index $i$ ranges from 1 to 4, $c_\beta({\mathbf p})$ is the annihilation operator for a certain species of particle and $d^\dagger_\beta({\mathbf p})$ is the creation operator for the corresponding antiparticle.  The index $\beta$ takes two values $\pm$.  The rest spinors are given up to proportionality by
\begin{equation}\label{gm:eqn:restelkos}
 \hspace*{-0.6em}\lambda^{\rm S}({\mathbf 0},+)=\left(\begin{array}{cccc}
 0\\
 i\\
 1\\
 0
 \end{array}\right), \lambda^{\rm S}({\mathbf 0},-)=\left(\begin{array}{cccc}
 i\\
 0\\
 0\\
 -1
 \end{array}\right), \lambda^{\rm A}({\mathbf 0},+)=\left(\begin{array}{cccc}
 0\\
 -i\\
 1\\
 0
 \end{array}\right), \lambda^{\rm A}({\mathbf 0},-)=\left(\begin{array}{cccc}
 -i\\
 0\\
 0\\
 -1
 \end{array}\right)
\end{equation}
and one obtains the spinors at nonzero momentum by multiplying the rest spinors by standard Lorentz boost matrices (Ref.~\refcite{gm:AhlGru05}, Sec.~3.3).

Recall the usual Dirac field \footnote{See, e.g., Ref.~\refcite{gm:Weinberg95}, Eq.~(5.5.34).  We have changed the signs of the exponential factors to be consistent with \eref{gm:eqn:elko}: this amounts simply to adopting a different convention in the definition of how the translation operator acts on physical states.}:
\begin{equation}\label{gm:eqn:dirac}
 \psi_i(x)= \int {\rm d}^3{\mathbf p} \sum_{\sigma}\left[e^{-ip_{\mu}x^{\mu}}u_i(\mathbf{p},\sigma)c_\sigma(\mathbf{p})+  e^{+ip_{\mu}x^{\mu}}v_i(\mathbf{p},\sigma)d^{\dagger}_\sigma(\mathbf{p})\right],
\end{equation}
where the rest spinors are given up to proportionality by
\begin{equation}\label{gm:eqn:restdirac}
 \hspace*{-1em}u({\mathbf 0},1/2)=\left(\begin{array}{cccc}
 1\\
 0\\
 1\\
 0
 \end{array}\right), u({\mathbf 0},-1/2)=\left(\begin{array}{cccc}
 0\\
 1\\
 0\\
 1
 \end{array}\right), v({\mathbf 0},1/2)=\left(\begin{array}{cccc}
 0\\
 1\\
 0\\
 -1
 \end{array}\right), v({\mathbf 0},-1/2)=\left(\begin{array}{cccc}
 -1\\
 0\\
 1\\
 0
 \end{array}\right).
\end{equation}
The Dirac rest spinors are eigenspinors of the helicity operator.  The starting point for the Elko construction is to choose rest spinors as in \eref{gm:eqn:restelkos} which are eigenspinors of the charge conjugation operator.  These eigenspinors are not eigenspinors of the helicity operator: the top two and bottom two components have opposite helicities.  For this reason Ahluwalia and Grumiller call $\beta$ a dual helicity index.

Having defined the Elko field, Ahluwalia and Grumiller introduce a dual --- which differs from the usual Dirac dual --- on the space of spinors.  They calculate the spin sums, the equation of motion and the propagator.  The propagator turns out to be the Klein-Gordan propagator and the mass dimension of the Elko field is 1 (as opposed to the value $3/2$ for the Dirac field).  This implies that the Elko field cannot interact with the electromagnetic field and that interactions --- ignoring gravity --- with all standard model particles, except possibly the Higgs boson, are prohibited or suppressed.  Hence the Elko field is a plausible candidate for a dark matter field.\footnote{The usual formalism of quantum field theory requires the fields to be local.  The original version of the Elko field in Ref.~\refcite{gm:AhlGru05} is not local.  Recently Ahluwalia, Lee, Schritt and Watson discovered a slightly different field which does satisfy locality\cite{gm:AhlLeeSch08}.  For simplicity we restrict ourselves to the original field in this note.}

\section{Weinberg's Definition of a Quantum Field}\label{gm:sec:weinberg}

In his book Ref.~\refcite{gm:Weinberg95}, Weinberg provides a broad and coherent framework for introductory field theory based on a few basic symmetry principles.  We briefly sketch an outline of his arguments and recall the relevant definitions.  See Ref.~\refcite{gm:Weinberg95}, Ch.~5, for more details.

The ingredients we need are the following.  We consider massive particles with positive energy, mass $m$ and spin $s$.  The little group ${\rm SO}(3)$ is a subgroup of ${\mathcal L}_0$.  Let $R(\Lambda)= R_{\sigma\nu}(\Lambda)$ be the $2s+1$-dimensional irreducible representation of ${\rm SO}(3)$ corresponding to spin $s$.  We construct a state space $H$ as in Ref.~\refcite{gm:Weinberg95}, Ch.~2: the space of one-particle states is spanned by basis kets of the form $\ket{p,\sigma}$, having 4-momentum $p$ and spin-$z$ component $\sigma$ in the rest frame.  Using the matrices $R_{\sigma\nu}(\Lambda)$, we can construct an irreducible unitary representation $U(\Lambda,a)$ of ${\mathcal P}_0$ on $H$.  Now let $D(\Lambda)= D_{ij}(\Lambda)$ be a $t$-dimensional representation of ${\mathcal L}_0$ for some $t\in {\mathbb N}$.  Let $L(H)$ denote the space of linear operators from $H$ to $H$.  We define a {\em Weinberg quantum field based on the data $(H,R(\Lambda),U(\Lambda,a),D(\Lambda))$} to be a collection of functions $\Psi(x)= (\Psi_i(x))_{1\leq i\leq t}$ from ${\mathbb R}^4$ to $L(H)$ such that for all $(\Lambda,a)\in {\mathcal P}_0$, we have
\begin{equation}\label{gm:eqn:fieldtransfn}
 U(\Lambda,a)\Psi_i(x)U(\Lambda,a)^{-1}= \sum_j D_{ij}(\Lambda^{-1}) \Psi_j(\Lambda x+ a).
\end{equation}

We say that a Weinberg quantum field --- or more generally a collection of Weinberg quantum fields --- is {\em local} if for any $\Psi$ and $\Phi$ in the collection, for any indices $i$ and $j$ and for any $x,y\in {\mathbb R}^4$ such that $x-y$ is spacelike, the field components $\Psi_i(x)$ and $\Phi_j(y)$ commute (for bosons) or anti-commute (for fermions).  Roughly speaking, \eref{gm:eqn:fieldtransfn} ensures that one can construct a Hamiltonian density ${\mathcal H}(x)$ from these fields which is a scalar under Poincar\'e transformations, and the extra requirement of locality ensures that the S-matrix obtained from ${\mathcal H}(x)$ transforms covariantly under Poincar\'e transformations.

It follows from quite general arguments that a Weinberg quantum field must be of the form\footnote{Note that for each fixed $x$, $i$, ${\mathbf p}$ and $\sigma$, all of the quantities that appear inside the integral in \eref{gm:eqn:stdform} except for $c_\sigma({\mathbf p})$ and $d^\dagger_\sigma({\mathbf p})$ are $c$-numbers.  Hence to evaluate the LHS of \eref{gm:eqn:fieldtransfn}, one need only calculate $U(\Lambda,a)c_\sigma({\mathbf p})U(\Lambda,a)^{-1}$ and $U(\Lambda,a)d^\dagger_\sigma({\mathbf p}) U(\Lambda,a)^{-1}$.}
\begin{equation}\label{gm:eqn:stdform}
 \Psi_i(x)= \int {\rm d}^3{\mathbf p}\sum_\sigma \left[e^{-ip_\mu x^\mu}u_i({\mathbf p},\sigma)c_\sigma({\mathbf p})+ e^{+ip_\mu x^\mu}v_i({\mathbf p},\sigma)d^\dagger_\sigma({\mathbf p})\right].
\end{equation}
\Eref{gm:eqn:fieldtransfn} implies that the spinors $u({\mathbf p},\sigma)$ and $v({\mathbf p},\sigma)$ for arbitrary ${\mathbf p}$ are completely determined by the rest spinors $u({\mathbf 0},\sigma)$ and $v({\mathbf 0},\sigma)$, and in many important cases, the values of $u({\mathbf 0},\sigma)$ and $v({\mathbf 0},\sigma)$ are also completely determined by \eref{gm:eqn:fieldtransfn} and the extra requirement of locality.

One advantage of Weinberg's approach is that physical insight falls out from the mathematical formalism: for instance, the Dirac equation --- and the form of the Dirac field itself --- can be derived from \eref{gm:eqn:fieldtransfn} and the requirement of locality (cf.~\sref{gm:sec:DiracvsElko}), so we may view them simply as consequences of covariance under the Poincar\'e group.  Similarly one can deduce the existence of anti-particles purely from the mathematical restrictions imposed by locality and \eref{gm:eqn:fieldtransfn} (see Ref.~\refcite{gm:Weinberg95}, Sec.~5.1).

\section{The Dirac and Elko Fields in Weinberg's Formalism}\label{gm:sec:DiracvsElko}

Throughout this section, we fix $D(\Lambda)= D_{ij}(\Lambda)$ to be the usual $(1/2,0)\oplus (0,1/2)$ representation of ${\mathcal L}_0$.  We wish to interpret the Elko field as a Weinberg quantum field $\Psi(x)$.  To do this, we must identify $\beta$ with the index $\sigma$ in the construction of $H$ above, for some suitable choice of $H$, $R(\Lambda)$ and $U(\Lambda,a)$.  Consider the state space $H$ and $U(\Lambda,a)$ constructed as above for $s=1/2$, where $R_{\sigma\nu}(\Lambda)$ is chosen to be the usual spin-$1/2$ representation of ${\rm SO}(3)$; the index $\sigma$ takes the values $\pm 1/2$.  Using \eref{gm:eqn:fieldtransfn} and the assumption of locality, one obtains formulas for the rest spinors which involve constants $c_{\pm}$, $d_{\pm}$. 
An argument involving parity conservation allows us to pin down the values of $c_{\pm}$ and $d_{\pm}$ up to an overall normalization (we return to this point in \sref{gm:sec:nonstdclasses} below).  We find that the rest spinors $u({\mathbf 0},\sigma)$ and $v({\mathbf 0},\sigma)$ are precisely the Dirac rest spinors given in \eref{gm:eqn:restdirac} above, and $\Psi(x)$ is the Dirac field (Ref.~\refcite{gm:Weinberg95}, Sec.~5.5).  Hence the Dirac field is the only Weinberg quantum field based on the data ($H$, $R(\Lambda)$, $U(\Lambda,a)$, $D(\Lambda)$).  This shows that the Elko field cannot be a Weinberg quantum field based on this data.  Indeed, Ahluwalia, Lee and Schritt have recently noted that the transformation properties of Elko spinors under rotations differ from the transformation rules that must be satisfied for a Weinberg spinor; the authors are grateful to them for communicating this observation to us.

We could instead have chosen the representation of ${\rm SO}(3)$ to be not the standard one $R(\Lambda)$ but another representation $R'(\Lambda)$ isomorphic to it.  Then $R(\Lambda)$ and $R'(\Lambda)$ are related by a similarity transform: we have $R'(\Lambda)= SR(\Lambda)S^{-1}$ for some invertible linear transformation $S$.  This change has no physical or mathematical significance, but it makes the resulting Weinberg quantum field look different.  One can show by adapting the argument on pp220--1 of Ref.~\refcite{gm:Weinberg95} that the rest spinors of a Weinberg quantum field based on $(H,R'(\Lambda),U(\Lambda,a),D(\Lambda))$ cannot take the form of the Elko rest spinors in \eref{gm:eqn:restelkos}, even if one does not assume parity conservation.  Hence the Elko field cannot be a Weinberg quantum field based on $(H,R'(\Lambda),U(\Lambda,a),D(\Lambda))$.  We will give full details of this calculation in forthcoming work.

\section{Non-standard Wigner Classes}\label{gm:sec:nonstdclasses}

As we have seen, the most direct attempt to fit the Elko field into Weinberg's formalism fails.  The next logical step, motivated by the discussion on p4, para.~1 of Ref.~\refcite{gm:AhlGru05}, is to consider Weinberg fields based on a state space $H$ and an irreducible representation of the full Poincar\'e group ${\mathcal P}$ on $H$ that belongs to one of the non-standard Wigner classes\cite{gm:Wigner64}.  Fix $R(\Lambda)$ and let $H$ be as in \sref{gm:sec:weinberg}, with one-particle basis kets $\ket{p,\sigma}$.  The unitary representation $U(\Lambda,a)$ of ${\mathcal P}_0$ extends to an (anti-)unitary representation $U(\Lambda,a)$ of the whole of ${\mathcal P}$; we abuse notation and denote this representation by $U(\Lambda,a)$ also.  The operators $U({\sf P})$ and $U({\sf T})$ are unitary and anti-unitary respectively, and they act on the kets $\ket{p,\sigma}$ by $U({\sf P})\ket{p,\sigma}= \eta_{\sf P}\ket{{\sf P}p,\sigma}$ and $U({\sf T})\ket{p,\sigma}= \eta_{\sf T}(-1)^{s-\sigma}\ket{{\sf P}p,-\sigma}$ for some constants $\eta_{\sf P}$ and $\eta_{\sf T}$ (Ref.~\refcite{gm:Weinberg95}, Sec.~2.6).  The representation $U(\Lambda,a)$ is said to belong to the standard Wigner class.

There are also three so-called non-standard Wigner classes of representations $U(\Lambda,a)$ of ${\mathcal P}$.  Here the states of given 4-momentum $p$ and spin-$z$ component $\sigma$ become degenerate; the basis kets are labelled $\ket{p,\sigma,\tau}$, where $\tau$ is an extra index that breaks the degeneracy.  Time reversal $U({\sf T})$ couples states with different values of $\tau$.  The Weinberg quantum fields in these degenerate cases are not worked out in detail in Ref.~\refcite{gm:Weinberg95} and we believe they are worth further study.  Even if we cannot give an interpretation of the Elko field in this setting, perhaps there are other as yet unexplored Weinberg quantum fields that may be candidates for dark matter.  Note that although the definition of a Weinberg field seems only to involve covariance under restricted Poincar\'e transformations, ${\sf P}$ and ${\sf T}$ play a crucial part (cf.~the final step in the derivation of the Dirac field in \sref{gm:sec:DiracvsElko}).

We finish with some remarks on work of Lee and Wick which is relevant here.  According to Ref.~\refcite{gm:LeeWick71}, if a field is local then the underlying representation of ${\mathcal P}$ must come from the standard Wigner class.  There, however, they allow themselves the freedom to multiply the original $U({\sf P})$ and $U({\sf T})$ by symmetries of the internal state space.  For the non-standard Wigner classes, where there are extra degrees of freedom coming from the index $\tau$, one would expect there to be plenty of these internal symmetries above and beyond charge conjugation.  A full study of the possible Weinberg fields would involve a systematic investigation of these internal symmetries.

\section*{Acknowledgments}

Most of the ideas in this note had their roots in discussions of the authors with Dharamvir Ahluwalia, Cheng-Yang Lee, Dimitri Schritt and Thomas Watson, and we thank them for their contribution and for their comments on an earlier draft.  The second author is grateful to Ahluwalia for introducing him to this area of research.

\bibliographystyle{ws-procs9x6}
\bibliography{gillard_martin3}

\end{document}